\documentclass{article}
\usepackage{kenmax,xspace,epsfig}
\def\bdoc{\begin{document}}
\def\edoc{\end{document}}

\def\babs{\begin{abstract}}
\def\eabs{\end{abstract}}

\def\bbib{\begin{thebibliography}{999}}
\def\ebib{\end{thebibliography}}

\def\beq{\begin{equation}}
\def\eeq{\end{equation}}

\def\beqn{\begin{eqnarray}}
\def\eeqn{\end{eqnarray}}

\def\beqnn{\begin{eqnarray*}}
\def\eeqnn{\end{eqnarray*}}

\def\barr{\begin{array}}
\def\earr{\end{array}}

\def\bqu{\begin{quote}}
\def\equ{\end{quote}}

\def\bqun{\begin{quotation}}
\def\equn{\end{quotation}}

\def\bit{\begin{itemize}}
\def\eit{\end{itemize}}

\def\ben{\begin{enumerate}}
\def\een{\end{enumerate}}

\def\bpic{\begin{picture}}
\def\epic{\end{picture}}

\def\bfig{\protect \begin{figure}}
\def\efig{\protect \end{figure}}

\def\bcc{\begin{center}}
\def\ecc{\end{center}}

\def\brr{\begin{flushright}}
\def\err{\end{flushright}}

\def\bll{\begin{flushleft}}
\def\ell{\end{flushleft}}

\def\btab{\begin{tabular}}
\def\etab{\end{tabular}}

\def\ul{\underline}
\def\ol{\overline}

\def\bm#1{\mbox{\boldmath $#1$}}

\def\zerob{{\bm 0}}
\def\oneb{{\bm 1}}

\def\ab{{\bm a}}
\def\bb{{\bm b}}
\def\cb{{\bm c}}
\def\db{{\bm d}}
\def\eb{{\bm e}}
\def\fb{{\bm f}}
\def\gb{{\bm g}}
\def\hb{{\bm h}}
\def\ib{{\bm i}}
\def\jb{{\bm j}}
\def\kb{{\bm k}}
\def\lb{{\bm l}}
\def\mb{{\bm m}}
\def\nb{{\bm n}}
\def\ob{{\bm o}}
\def\pb{{\bm p}}
\def\qb{{\bm q}}
\def\rb{{\bm r}}
\def\sb{{\bm s}}
\def\tb{{\bm t}}
\def\ub{{\bm u}}
\def\vb{{\bm v}}
\def\wb{{\bm w}}
\def\xb{{\bm x}}
\def\yb{{\bm y}}
\def\zb{{\bm z}}

\def\Ab{{\bm A}}
\def\Bb{{\bm B}}
\def\Cb{{\bm C}}
\def\Db{{\bm D}}
\def\Eb{{\bm E}}
\def\Fb{{\bm F}}
\def\Gb{{\bm G}}
\def\Hb{{\bm H}}
\def\Ib{{\bm I}}
\def\Jb{{\bm J}}
\def\Kb{{\bm K}}
\def\Lb{{\bm L}}
\def\Mb{{\bm M}}
\def\Nb{{\bm N}}
\def\Ob{{\bm O}}
\def\Pb{{\bm P}}
\def\Qb{{\bm Q}}
\def\Rb{{\bm R}}
\def\Sb{{\bm S}}
\def\Tb{{\bm T}}
\def\Ub{{\bm U}}
\def\Vb{{\bm V}}
\def\Wb{{\bm W}}
\def\Xb{{\bm X}}
\def\Yb{{\bm Y}}
\def\Zb{{\bm Z}}

\def\alphab{\bm{\alpha}}
\def\betab{\bm{\beta}}
\def\deltab{\bm{\delta}}
\def\epsilonb{\bm{\epsilon}}
\def\gammab{\bm{\gamma}}
\def\omegab{\bm{\omega}}
\def\thetab{\bm{\theta}}
\def\xib{\bm{\xi}}
\def\lambdab{\bm{\lambda}}
\def\taub{\bm{\tau}}
\def\phib{\bm{\phi}}
\def\mub{\bm{\mu}}
\def\psib{\bm{\psi}}
\def\chib{\bm{\chi}}
\def\sigmab{\bm{\sigma}}

\def\Deltab{\bm{\Delta}}
\def\Lambdab{\bm{\Lambda}}
\def\Phib{\bm{\Phi}}
\def\Psib{\bm{\Psi}}
\def\Sigmab{\bm{\Sigma}}

\def\Ac{{\cal A}}
\def\Bc{{\cal B}}
\def\Cc{{\cal C}}
\def\Dc{{\cal D}}
\def\Ec{{\cal E}}
\def\Fc{{\cal F}}
\def\Gc{{\cal G}}
\def\Hc{{\cal H}}
\def\Ic{{\cal I}}
\def\Jc{{\cal J}}
\def\Kc{{\cal K}}
\def\Lc{{\cal L}}
\def\Mc{{\cal M}}
\def\Nc{{\cal N}}
\def\Oc{{\cal O}}
\def\Pc{{\cal P}}
\def\Qc{{\cal Q}}
\def\Rc{{\cal R}}
\def\Sc{{\cal S}}
\def\Tc{{\cal T}}
\def\Uc{{\cal U}}
\def\Vc{{\cal V}}
\def\Wc{{\cal W}}
\def\Xc{{\cal X}}
\def\Yc{{\cal Y}}
\def\Zc{{\cal Z}}

\def\wt#1{\widetilde{#1}}
\def\wh#1{\widehat{#1}}
\def\xh{\widehat{x}}
\def\thetah{\widehat{\theta}}
\def\betah{\widehat{\beta}}

\def\xbh{\widehat{\xb}}
\def\thetabh{\widehat{\thetab}}
\def\betabh{\widehat{\betab}}

\def\xbhk{\widehat{\xb}^{k}}
\def\thetahk{\widehat{\theta}^{k}}
\def\betahk{\widehat{\beta}^{k}}

\def\xbhkp{\widehat{\xb}^{k+1}}
\def\thetahkp{\widehat{\theta}^{k+1}}
\def\betahkp{\widehat{\beta}^{k+1}}

\def\thetabhk{\widehat{\thetab}^{k}}
\def\betabhk{\widehat{\betab}^{k}}

\def\thetabhkp{\widehat{\thetab}^{k+1}}
\def\betabhkp{\widehat{\betab}^{k+1}}

\def\thetamin{\theta_{\mbox{\tiny min}}}
\def\thetamax{\theta_{\mbox{\tiny max}}}
\def\betamin{\beta_{\mbox{\tiny min}}}
\def\betamax{\beta_{\mbox{\tiny max}}}

\def\ra{\rightarrow}
\def\la{\leftarrow}
\def\da{\downarrow}
\def\ua{\uparrow}

\def\Ra{\Rightarrow}
\def\La{\Leftarrow}
\def\Da{\Downarrow}
\def\Ua{\Uparrow}

\def\lra{\longrightarrow}
\def\lla{\longleftarrow}
\def\Lra{\Longrightarrow}
\def\Lla{\longleftarrow}

\def\lrarr{\leftrightarrow}
\def\Lrarr{\Leftrightarrow}
\def\udarr{\updownarrow}
\def\Uparr{\Updownarrow}

\def\d#1{\,\mbox{d}#1}
\def\dxdy{\d{x}\d{y}}
\def\dwxdwy{\d{\omega_x}\d{\omega_y}}
\def\dxdydz{\d{x}\d{y}\d{z}}

\def\disp#1{{\displaystyle #1}}
\def\diag#1{\mbox{diag}\left\{#1\right\}}

\def\Prob#1{\mbox{Pr}\left\{#1\right\}}
\def\var#1{\mbox{Var}\left\{#1\right\}}
\def\cov#1{\mbox{Cov}\left\{#1\right\}}
\def\corr#1{\mbox{Corr}\left\{#1\right\}}
\def\trace#1{\mbox{Tr}\left\{#1\right\}}
\def\rang#1{\mbox{rang}\left\{#1\right\}}
\def\det#1{\mbox{d\'et}\left\{#1\right\}}

\def\cosf{\cos \phi}
\def\sinf{\sin \phi}
\def\cost{\cos \theta}
\def\sint{\sin \theta}

\def\sgn{\mbox{sgn}}
\def\sinc{\mbox{sinc}}
\def\rect{\mbox{rect}}
\def\sincf#1{\mbox{sinc}\left(#1\right)}
\def\rectf#1{\mbox{rect}\left(#1\right)}
\def\trif#1{\mbox{tri}\left(#1\right)}
\def\xvec#1#2#3{\left\{#1_#2,\ldots,#1_#3\right\}}

\def\vx{\left[x_1,\ldots, x_n\right]^t}
\def\vz{\left[z_1,\ldots, z_n\right]^t}
\def\vw{\left[\omega_1,\ldots, \omega_n\right]^t}
\def\vxi{\left[\xi_1,\ldots, \xi_n\right]^t}
\def\iii{\int_{-\infty}^{+\infty}}
\def\izi{\int_{0}^{\infty}}
\def\izpi{\int_{0}^{\pi}}
\def\izdpi{\int_{0}^{2\pi}}
\def\intd{\int\kern-.8em\int}
\def\intt{\int\kern-.8em\int\kern-.8em\int}
\def\intg{\int\kern-1.1em\int}
\def\sumd{\mathop{\sum\sum}}

\def\sumi{\sum_{i=1}^{M}}
\def\sumj{\sum_{i=1}^{N}}
\def\sumk{\sum_{k=1}^{K}}
\def\sumn{\sum_{n=1}^{N}}
\def\summ{\sum_{m=1}^{M}}

\def\TA#1{{\cal A}\left\{ {#1} \right\}}
\def\TH#1{{\cal H}\left\{ {#1} \right\}}
\def\TP#1{{\cal P}\left\{ {#1} \right\}}
\def\TR#1{{\cal R}\left\{ {#1} \right\}}
\def\TRa#1{{\cal R}^{\dag}\left\{ {#1} \right\}}
\def\BR#1{{\cal B}\left\{ {#1} \right\}}
\def\TF#1{{\cal F}\left\{ {#1} \right\}}
\def\TFI#1{{\cal F}^{-1}\left\{ {#1} \right\}}
\def\TFn#1#2{{\cal F}_{#1}\left\{ {#2} \right\}}
\def\TFnI#1#2{{\cal F}_{#1}^{-1}\left\{ {#2} \right\}}
\def\Im#1{{\cal I}\mbox{m}\left(#1\right)}
\def\Ker#1{{\cal K}\mbox{er}\left(#1\right)}
\def\Imag#1{\mbox{Im}\left(#1\right)}
\def\Re#1{\mbox{Re}\left(#1\right)}
\def\expf#1{\exp\left[ {#1} \right]}

\def\dfdx#1#2{{\mbox{d} {#1}\over{\mbox{d} {#2}}}}
\def\dfdxd#1#2{{\mbox{d}^2 {#1}\over{\mbox{d} {#2}^2}}}
\def\dfdxt#1#2{{\mbox{d}^3 {#1}\over{\mbox{d} {#2}^3}}}
\def\dfdxn#1#2{{\mbox{d}^n {#1}\over{\mbox{d} {#2}^n}}}
\def\dfdxk#1#2{{\mbox{d}^k {#1}\over{\mbox{d} {#2}^k}}}

\def\dpdx#1#2{{{\partial {#1}\over \partial {#2}}}}
\def\dpdxd#1#2{{{\partial^2 {#1}}\over{\partial {#2}^2}}}
\def\dpdxdy#1#2#3{{{\partial ^2 {#1}}\over{\partial {#2} \partial {#3}}}}

\def\arg{\mbox{arg}}
\def\argmins#1#2{\mbox{arg}\min_{#1}\left\{{#2}\right\}}
\def\argmaxs#1#2{\mbox{arg}\max_{#1}\left\{{#2}\right\}}
\def\argmin#1#2{\mathop{\mbox{arg}\min}_{#1}\left\{{#2}\right\}}
\def\argmax#1#2{\mathop{\mbox{arg}\max}_{#1}\left\{{#2}\right\}}

\def\esp#1{\mbox{E}\left\{ #1 \right\}}
\def\espx#1#2{\mbox{E}_{#1}\left\{ #2 \right\}}

\def\wth#1{\widehat{\widetilde{\phantom{#1}}}\!\!\!\! #1}

\def\lrf{L_{r,\phi}}
\def\fw{\widehat{f}(\omegab)}
\def\fthwxi{\wth{f}(\Omega,\xib)}
\def\fthwfi{\wth{f}(\Omega,\phi)}
\def\ftrfi{\widetilde{f}(r,\phi)}

\def\fwxwy{\widehat{f}(\omega_x, \omega_y)}
\def\wxpwy{(\omega_x \, x + \omega_y \, y)}

\def\wtx{\omegab^t \cdot \xb}
\def\ejwtx{\exp\left[j \omegab^t \cdot \xb\right]}
\def\xitx{\xib^t \cdot \xb}

\def\ftrxi{\widetilde{f}(r,\xib)}
\def\ent{-\int p(x) \, \ln p(x) \d{x}}

\def\mean#1{\left< #1 \right>}
\def\slnhn{\sum_{n=1}^N \lambda_n h_n(\rb)}
\def\slngn{\sum_{n=1}^N \lambda_n g_n(\rb)}
\def\smngn{\sum_{n=1}^N \mu_n g_n(\rb)}
\def\slmhm{\sum_{m=1}^N \lambda_m h_m(\rb)}
\def\vlambda{\bm{\lambda} = [\lambda_1,\ldots,\lambda_n]}

\def\apriori{{\em a priori} }
\def\aposteriori{{\em a posteriori} }

\def\titre#1{\bcc{\Large\bf #1}\ecc}

\def\AMD{Ali Mohammad--Djafari}
\def\LSSa{Laboratoire des Signaux et Syst\`emes 
(CNRS--ESE--UPS) \\ 
\'Ecole Sup\'erieure d'\'Electricit\'e \\ 
Plateau de Moulon, 91192 Gif sur Yvette Cedex, France.}

\def\ME{maximum entropy}
\def\pdf{probability distribution function}
\def\lm{Lagrange multipliers}
\def\fix#1{\phi _#1(x)}
\def\fin{\fix n}
\def\fik{\fix k}
\def\fiz{\fix 0}
\def\sfinz{\sum_{n=0}^N \lambda_n \, \fin}
\def\sfinu{\sum_{n=1}^N \lambda_n \, \fin}
\def\bl{\bm{\lambda}}
\def\bd{\bm{\delta}}
\def\blz{\bl ^0}
\def\gnl{G _n(\bl)}
\def\gnlz{G _n(\blz)}
\def\un{n=1,\dots, N}
\def\nn{n=0,\dots, N}

\def\finn{\fin , \nn}
\def\esfinz{\exp\,\left[ -\sfinz \right] }
\def\esfinu{\exp\,\left[ -\sfinu \right] }
\def\esxm{\exp\,\left[ -\sum_{m=0}^N \lambda_m \, x^m \right] }
\def\efin{\esp \fin }
\def\zl{Z(\bl)}
\def\finxi{\phi _n(x_i)}
\def\snfinxi{\sum_{n=1}^N \lambda_n \finxi}
\def\esnfinxi{\exp \left[ - \snfinxi \right]}
\def\smfinxi{\sum_{i=1}^M \finxi}

\def\ejnw{\exp \left( -j n \omega_0 x \right) }
\def\eejnw{\mbox{E} \left\lbrace \ejnw \right\rbrace}

\def\signed#1{{\unskip\nobreak\hfil\penalty50\hskip2em\mbox{}
\nobreak\hfil\tt#1\parfillskip=0pt \finalhyphendemerits=0 \par}}

\def\uncatcodespecials{\def\do##1{\catcode`##1=12 }\dospecials}
\def\listing#1{\par\begingroup\setupverbatim\input#1 \endgroup}
\newcount\lineno
\def\setupverbatim{\tt \lineno=0
 \obeylines \uncatcodespecials \obeyspaces
 \everypar{\advance\lineno by1 \llap{\sevenrm\the\lineno\ \ }}}
{\obeyspaces\global\let =\ }

\def\defined{\stackrel{\mbox{def}}{=}}
\def\str{\stackrel}

\def\ER{\mbox{I\kern-.25em R}}
\def\EC{\mbox{C\kern-.8em C}}
\def\EZ{\mbox{Z\kern-.55em Z}}
\def\EN{\mbox{N\kern-.8em N}}

\def\singles{
 \abovedisplayskip 12pt plus 3pt minus 9pt
 \belowdisplayskip 12pt plus 3pt minus 9pt
 \abovedisplayshortskip 0pt plus 3pt
 \belowdisplayshortskip 7pt plus 3pt minus 4pt
 \baselineskip 14.4pt
 \lineskip 1pt
 \lineskiplimit 0pt}
\def\oneandhalf{
 \abovedisplayskip 18pt plus 3pt minus 9pt
 \belowdisplayskip 18pt plus 3pt minus 9pt
 \abovedisplayshortskip 0pt plus 3pt
 \belowdisplayshortskip 9.333pt plus 3pt
 \baselineskip 20pt
 \lineskip 2pt
 \lineskiplimit 1pt}

\def\double{
 \abovedisplayskip 24pt plus 3pt minus 9pt
 \belowdisplayskip 24pt plus 3pt minus 9pt
 \abovedisplayshortskip 0pt plus 3pt
 \belowdisplayshortskip 12pt plus 3pt
 \baselineskip 27pt
 \lineskip 3pt
 \lineskiplimit 2pt}

\def\dadb{\d{\alpha}\d{\beta}}

\def\ffbox#1{\fbox{\mbox{\vbox{#1}}}}

\def\rot{\mbox{rot}}
\def\case#1#2#3#4{
    \left\{
           \begin{array}{ll}
            {\displaystyle #1} & {\displaystyle #2} \cr 
            {\displaystyle #3} & {\displaystyle #4}
           \end{array}
    \right. }

\def\beqnarr#1&#2&#3\\#4&#5&#6\eeqnarr{
    \left\{
           \begin{array}{lcl}
            {\displaystyle #1} & #2 & {\displaystyle #3} \\ 
            {\displaystyle #4} & #5 & {\displaystyle #6} 
           \end{array}
    \right. }

\def\pyx{p(\yb|\xb)}
\def\pxy{p(\xb|\yb)}

\def\ie{{\em i.e.}}
\def\unsdpi{\left(\frac{1}{2\pi}\right)}
\def\unspi{\left(\frac{1}{\pi}\right)}
\def\up{\uppercase}
\def\zjm{z_{j-1}}
\def\zjp{z_{j+1}}
\def\fxyp{f(x,y)=\left\{
\barr{ll} 1 & (x,y)\in P\\ 0 & (x,y)\not\in P\earr
\right.}

\def\syslins#1#2#3#4{
\bpic(150,30)
  \put(5,5){\makebox(15,0){$#1$}}
  \put(20,5){\vector(1,0){10}}
  \put(30,0){\framebox(30,10){$#2$}}
  \put(60,5){\vector(1,0){10}}
  \put(75,5){\circle{10}}
  \put(75,5){\makebox(0,0){$+$}}
  \put(75,20){\vector(0,-1){10}}
  \put(68,22){\makebox(15,0){$#3$}}
  \put(80,5){\vector(1,0){10}}
  \put(100,5){\makebox(50,0){$#4=#2 #1 + #3$}}
\epic
}
\def\siso#1#2#3{
#1\mapsto\fbox{#2}\mapsto#3
}
\def\sido#1#2#3#4{
#1\mapsto 
\fbox{\small\begin{tabular}{c} ~ \\ {#2} \\ ~ \end{tabular}}  
\begin{array}{l} \mapsto #3 \\ \mapsto #4\end{array}
}
\def\sidotp#1#2#3#4#5#6{
#1\mapsto 
\fbox{\small\begin{tabular}{c} {#2} \\ {#3} \\ {#4} \end{tabular}}  
\begin{array}{l} \mapsto #5 \\ \mapsto #6\end{array}
}
\def\sidotpup#1#2#3#4#5#6{
#1\mapsto 
\fbox{\small\begin{tabular}{c} {#2} \\ {#3} \\ {#4} \end{tabular}}  
\begin{array}{l} \mapsto #5 \\ \mapsto #6\\ ~ \end{array}
}
\def\sidodn#1#2#3#4{
#1\mapsto 
\fbox{\small\begin{tabular}{c} ~ \\ {#2} \\ ~ \end{tabular}}  
\begin{array}{l} ~\\ \mapsto #3 \\ \mapsto #4\end{array}
}
\def\sidoup#1#2#3#4{
#1\mapsto 
\fbox{\small\begin{tabular}{c} ~ \\ {#2} \\ ~ \end{tabular}}  
\begin{array}{l} \mapsto #3 \\ \mapsto #4 \\ ~ \end{array}
}
\def\diso#1#2#3#4{
\begin{array}{r}  #1\mapsto \\ #2\mapsto \end{array}
\fbox{\small\begin{tabular}{c}~\\ {#3} \\~\end{tabular}}\mapsto#4
}
\def\disod#1#2#3#4{
\begin{array}[t]{c}#2\\ \downarrow \\ #1\mapsto\fbox{#3}\mapsto#4\end{array} 
}
\def\disou#1#2#3#4{
\begin{array}[t]{c}#1\mapsto\fbox{#3}\mapsto #4 \\ \uparrow \\ #2\end{array} 
}
\def\dido#1#2#3#4#5{
\begin{array}{r}  #1\mapsto \\ #2\mapsto \end{array}
\fbox{\small\begin{tabular}{c} ~ \\ {#3} \\ ~ \end{tabular}}  
\begin{array}{l} \mapsto #4 \\ \mapsto #5\end{array}
}
\def\didou#1#2#3#4#5{
#1\mapsto
\begin{array}{c} 
 #2 \\ \downarrow \\ 
 \fbox{\small\begin{tabular}{c} ~ \\ {#3} \\ ~ \end{tabular}} \\ 
 ~ \\ ~  
\end{array}
\begin{array}{l}  \mapsto #4 \\ \mapsto #5\end{array}
}
\def\didod#1#2#3#4#5{
#1\mapsto
\begin{array}{c} 
 ~ \\ ~\\ 
 \fbox{\small\begin{tabular}{c} ~ \\ {#3} \\ ~ \end{tabular}} \\ 
 \uparrow \\ #2 
\end{array}
\begin{array}{l}  \mapsto #4 \\ \mapsto #5\end{array}
}
\def\dito#1#2#3#4#5#6{
\begin{array}{r}  #1\mapsto \\ #2\mapsto \end{array}
\fbox{\small\begin{tabular}{c} ~ \\ {#3} \\ ~ \end{tabular}}  
\begin{array}{l} \mapsto #4 \\ \mapsto #5 \\ \mapsto #6\end{array}
}
\def\tido#1#2#3#4#5#6{
\begin{array}{r}  #1\mapsto\\ #2\mapsto\\ #3\mapsto\end{array}
\fbox{\small\begin{tabular}{c} ~ \\ {#4} \\ ~ \end{tabular}}
\begin{array}{l}  \mapsto #5 \\ \mapsto #6\end{array}
}
\def\tidoud#1#2#3#4#5#6{
#1\mapsto
\begin{array}{c} 
 #2 \\ \downarrow \\ 
 \fbox{\small\begin{tabular}{c} ~ \\ {#4} \\ ~ \end{tabular}} \\ 
 \uparrow \\ #3 
\end{array}
\begin{array}{l}  \mapsto #5 \\ \mapsto #6\end{array}
}
\def\tiso#1#2#3#4#5{
\begin{array}{r}  #1\mapsto\\ #2\mapsto\\ #3\mapsto\end{array}
\fbox{\small\begin{tabular}{c} ~ \\ {#4} \\ ~ \end{tabular}}
\mapsto #5
}
\def\tisoud#1#2#3#4#5{
#1\mapsto
\begin{array}{c} 
 #2 \\ \downarrow \\ 
 \fbox{\small\begin{tabular}{c} ~ \\ {#4} \\ ~ \end{tabular}} \\ 
 \uparrow \\ #3 
\end{array}
\mapsto #5
}

\title{Probabilistic methods for data fusion}

\author{Ali Mohammad--Djafari\\[12pt]
Laboratoire des Signaux et Syst\`emes (CNRS--SUPELEC--UPS) \\
\'Ecole Sup\'erieure d'\'Electricit\'e \\
Plateau de Moulon, 91192 Gif--sur--Yvette Cedex, France. \\
E\_mail: {\tt djafari@lss.supelec.fr}
}
\date{}

\begin{document}
\maketitle

\index{Probabilistic methods}
\index{Data fusion}
\index{Maximum likelihood}
\index{Maximum entropy}
\index{EM algorithm}

\begin{abstract}
The main object of this paper is to show how we can use classical
probabilistic methods such as
Maximum Entropy (ME), maximum likelihood (ML) and/or Bayesian (BAYES)
approaches to do microscopic and macroscopic data fusion.
Actually ME can be used to assign a probability law to an unknown quantity
when we have macroscopic data (expectations) on it.
ML can be used to estimate the parameters of a probability law
when we have microscopic data (direct observation).
BAYES can be used to update a prior probability law when we have
microscopic data through the likelihood.
When we have both microscopic and macroscopic data we can use first
ME to assign a prior and then use BAYES to update it to the posterior
law thus doing the desired data fusion.
However, in practical data fusion applications, we may still need some
engineering feeling to propose realistic data fusion solutions.
Some simple examples in sensor data fusion and image reconstruction using
different kind of data are presented to illustrate these ideas.
\\ ~\\
{\bf key words:}~ Data fusion, Maximum entropy, Maximum likelihood,
Bayesian data fusion, EM algorithm.
\end{abstract}

\section{Introduction}

Data fusion is one of the active area of research in many applications
such as non destructive testing (NDT), geophysical imaging,
medical imaging, radio-astronomy, etc.
Our main object in this paper is not to focus on any of these applications.
Indeed, we want to show how we can use classical
probabilistic methods such as
Maximum Entropy (ME), maximum likelihood (ML) and/or Bayesian (BAYES)
approaches to do data fusion.

First, we consider these three methods separately, and
we describe briefly each method.
Then we will see some interrelations between them.

We will see that ME can be used to assign a probability law to an
unknown quantity $X$ when we have macroscopic data (expectations) on it.
ML can be used when we have assigned a parametric probability law,
before getting the data, on $X$ and
we want to estimate this parameter from some microscopic data
(samples of $X$).
BAYES can be used to update probability laws,
going from priors to posteriors.

When we have both microscopic and macroscopic data we can use first
ME to assign a prior and then use BAYES to update it to the posterior
law, doing thus the desired data fusion.
In practical data fusion applications, however, we may still need some
engineering feeling to propose realistic data fusion solutions.

\section{Short description of the methods}

\subsection{Maximum Entropy (ME)}

ME can be used to assign a probability law to an
unknown quantity when we have macroscopic data (expectations) on it.
To see this let note by $X$ a quantity of interest and try to see when
and how we can use ME. We do this through a given problem.

\medskip\noindent{\em Problem P1:}
We have $L$ sensors giving us $L$ values $\{\mu_l, l=1,\ldots,L\}$,
representing the mean values of $L$ known functions
$\left\{ \phi_l(X), l=1,\ldots,L \right\}$ related to the unknown $X$:
\beq
\esp{\phi_l(X)}=\int \phi_l(x) p(x) \d{x}=\mu_l,\quad l=1,\ldots,L.
\eeq
The question is then how to represent our partial knowledge of
$X$ by a probability law.

Obviously, this problem has not a unique solution. Actually these data
define a class of possible solutions and we need a criterion to select
one of them. The ME principle can give us this criterion and the problem
then becomes:

\[
\hbox{maximize}\qquad\qquad
S(p)=-\int p(x) \, \ln p(x) \d{x}
\]
\[
\hbox{subject to}\qquad\qquad
\int \phi_l(x) \, p(x) \d{x}=\mu_l ,\quad l=1,\ldots,L.
\]
The solution is given by
\beq
p(x)=\frac{1}{Z(\thetab)} \, \expf{- \sum_{l=1}^L \theta_l \phi_l(x)}
     =\frac{1}{Z(\thetab)} \, \expf{- \thetab^t \phib(x)},
\eeq
where
\beq
 Z(\thetab)=\int \expf{-\sum_{l=1}^L \theta_l \phi_l(x)} \d{x}
\eeq
is the partition function and
$\{\theta_1,\ldots,\theta_l\}$ are determined by the following system of equations:
\begin{equation} \label{solME}
\fbox{$\displaystyle -\frac{\partial \ln Z(\thetab)}{\partial\theta_l}=\mu_l,
 \quad l=1,\ldots,L,$}
\end{equation}
See \cite{Djafari92,Djafari94} for more discussions.

\bigskip
\subsection{Maximum Likelihood (ML)}

\medskip\noindent{\em Problem P2:}
Assume now that we have a parametric form of the probability law
$p(x;\thetab)$ and a sensor gives us $N$ values
$\bm{x}=[x_1, \ldots, x_N]$ of $X$.
How to determine the parameters $\thetab$?

Two classical methods for solving this problem are:
\begin{itemize}
\item Moments Method (MM): The main idea is to write a set of equations
(at least $L$) relating the theoretical and empirical moments,
and solve them to obtain the solution:
\beq
G_l(\thetab)=\esp{X^l}=\int x^l \, p(x;\thetab) \d{x}
                   =\frac{1}{N} \sum_{j=1}^N x_j^l, \, l=1,\ldots,L
\eeq
\item Maximum Likelihood (ML): Here, the main idea is to consider the data
as $N$ samples of $X$.
Then, writing the expression of $p(\bm{x}; \thetab)$ and considering it as a
function of $\thetab$, the ML solution is defined as
\beq
\wh{\thetab}=\argmaxs{\thetab}{l(\thetab|\bm{x})}
\hbox{~with~}
l(\thetab|\bm{x}) =p(\bm{x}; \thetab)=\prod_{j=1}^N p(x_j; \thetab)
\eeq
\end{itemize}

It is interesting to note that, in the case of the generalized
exponential families:
\beq \label{gexp}
p(x;\thetab)
=\frac{1}{Z(\thetab)} \expf{-\sum_{l=1}^L \theta_l \, \phi_l(x)}
=\frac{1}{Z(\thetab)} \expf{-\thetab^t \, \phib(x)}
\eeq
we have
\beq
l(\thetab)=\prod_{j=1}^N p(x_j; \thetab)
=\frac{1}{Z^N(\thetab) }
\expf{- \sum_{j=1}^N \sum_{l=1}^L \theta_l \,\phi_l(x_j)}
\eeq
Then, it is easy to see that the ML solution is the solution of the following
system of equations:
\begin{equation} \label{solMV}
\fbox{$\displaystyle \dpdx{\ln Z(\thetab)}{\theta_l}
=\frac{1}{N} \sum_{j=1}^N \phi_l(x_j),\quad l=1,\ldots,L$}
\end{equation}
Comparing equations (\ref{solME}) \& (\ref{solMV}), we can remark an
interesting relation between these two methods.
See also \cite{Djafari91} for more discussions.

\bigskip
\subsection{ML and incomplete data: EM Algorithm}

\medskip\noindent{\em Problem P3:}
Consider the {\em  problem P2}, but now assume that the sensor gives
us $M$ values
$\yb=[y_1, \ldots, y_M]$ related to the $N$ samples
$\xb=[x_1, \ldots, x_N]$ of $X$ by a non invertible relation,
$\bm{y}=\bm{A} \bm{x}$ with $M<N$.
How to determine $\thetab$~?

The solution here is still based on the ML. The only difference is the way
to calculate the solution. In fact we can write
\beq
p(\xb;\thetab)=p(\xb|\yb;\thetab) \, p(\yb;\thetab),
\qquad \forall \Ab \xb=\yb.
\eeq
Taking the expectation of both sides for a given value of $\thetab=thetab'$,
we have
\beq
\ln p(\yb;\thetab)=\espx{\xb|\yb;\thetab'}{\ln p(\xb;\thetab)}
                     -\espx{\xb|\yb;\thetab'}{\ln p(\xb|\yb;\thetab)}
\eeq
or written differently:
\beq
L(\thetab)=Q(\thetab;\thetab')-V(\thetab,\thetab').
\eeq
Note that for a given $\thetab'$ and for all $\thetab$ we have
\beq
L(\thetab)-L(\thetab')
= [Q(\thetab;\thetab')-Q(\thetab';\thetab')] +
   [V(\thetab;\thetab')-V(\thetab',\thetab')].
\eeq
Now, using the Jensen's inequality \cite{Miller87}
\beq
V(\thetab;\thetab') \le V(\thetab',\thetab')
\eeq
an iterative algorithm, known as {\em Expectation-Maximization (EM)},
is derived:
\beq
\left\{\begin{array}{lll}
\hbox{E:~~} &Q\left(\thetab;\wh{\thetab}^{(k)}\right)
       &= \disp{ \espx{\xb|\yb;\thetab^{(k)}}{\ln p(\xb;\thetab)} } \\
\hbox{M:~~} &\wh{\thetab}^{(k+1)}
       &= \disp{ \argmaxs{\thetab}{Q\left(\thetab;\wh{\thetab}^{(k)}\right)} }
\end{array}
\right.
\eeq
This algorithm insures to converge to a local maximum of the likelihood.

It is interesting to see that in the case of the generalized exponential
families (\ref{gexp}), the algorithm becomes: \\ ~\\
\beqn
\hbox{Step E:} &&
Q(\thetab;\thetab')
=\espx{\xb|\yb;\thetab'}{\ln p(\xb;\thetab)}
=-N \ln Z(\thetab)-\sum_{j=1}^N \thetab^t \espx{\xb|\yb;\thetab'}{\phib(x_j)}
\nonumber \\
\hbox{Step M:} &&
\fbox{$\displaystyle-\dpdx{\ln Z(\thetab)}{\theta_l}
=\frac{1}{N} \sum_{j=1}^N \espx{x_j|\yb;\thetab^{(k)}}{\phi_l(x_j)},
\quad l=1,\ldots,L$}  \label{solEM}
\eeqn
Compare this last equation with those of (\ref{solME}) and (\ref{solMV})
to see still some relations between ME, ML and the EM algorithms.

\medskip\noindent{\em Problem P4:}
Consider now the same problem {\em P3} where we want to estimate
not only $\thetab$ but also $\xb$.
We can still use the EM algorithm with the following modification:
\beq
\left\{\begin{array}{lll}
\hbox{E:~~}
 & Q\left(\thetab;\wh{\thetab}^{(k)}\right)
 &=\esp{ \ln p(\xb;\thetab)|\yb;\wh{\thetab}^{(k)} }
\\
 & \wh{\xb}^{(k)}
 &=\disp{ \esp{\xb|\yb;\wh{\thetab}^{(k)}} }
\\
\hbox{M:~~}
 & \wh{\thetab}^{(k+1)}
 &= \disp{ \argmaxs{\thetab}{Q\left(\thetab;\wh{\thetab}^{(k)}\right)} }
\end{array}\right.
\eeq

\subsection{Bayesian Approach}

\medskip\noindent{\em  Problem P5:}
Consider again {\em problems P3} or {\em P4} but now assume that the
observations $\bm{y}$ are corrupted by noise: $\bm{y}=\bm{A} \bm{x}+\bm{b}$.

The main tool here is the Bayesian approach where, we use the data-unknown
relation and the noise probability distribution to define the likelihood
$p(\yb|\xb;\thetab_1)=p_b(\yb-\Ab \xb;\thetab_1)$ and combine it with the
prior $p(\xb;\thetab_2)$ through the Bayes' rule to obtain the posterior
law
\beq
p(\xb|\yb;\thetab_1,\thetab_2)
=\frac{p(\yb|\xb;\thetab_1)\, p(\xb;\thetab_2)}{m(\yb;\thetab_1,\thetab_2)},
\eeq
where
\beq
m(\yb;\thetab_1,\thetab_2)=\intg p(\yb|\xb;\thetab_1)\, p(\xb;\thetab_2) \d{\xb}
\eeq
The posterior law $p(\xb|\yb;\thetab_1,\thetab_2)$ contains all the
information available on $\xb$. We can then use it to make any inference
on $\xb$. We can for example define the following point estimators:
\bit
\item Maximum \aposteriori (MAP):
\beq
\wh{\xb}=\argmaxs{\xb}{p_{x|y}(\xb|\yb;\thetab_1,\thetab_2)}
\eeq
\item Posterior Mean (PM):
\beq
\wh{\xb}=\espx{x|y}{\xb}
=\intg \xb \, p_{x|y}(\xb|\yb;\thetab_1,\thetab_2) \d{\xb}
\eeq
\item Marginal Posterior Modes (MPM):
\beq
\wh{\xb}=\argmaxs{x_i}{p(x_i|\yb;\thetab_1,\thetab_2)},
\eeq
where
\beq
p(x_i|\yb;\thetab_1,\thetab_2)=\intg p_{x|y}(\xb|\yb;\thetab)
 \d{x}_1 \ldots \d{x}_{i-1} \ldots \d{x}_{i+1} \ldots \d{x}_n
\eeq
\eit
However, in practice, we face two great difficulties:
\bit
\item How to assign the probability laws $p(\yb|\xb;\thetab_1)$
and $p(\xb;\thetab_2)$?
\item How to determine the parameters $\thetab=(\thetab_1,\thetab_2)$?
\eit
For the first we can use either the ME principle when possible,
or any other invariance properties combined with some practical,
scientific or engineering sense reasoning. For the second,
there are more specific tools,
all based on the joint posterior probability law
\[
p(\xb,\thetab|\yb)
\propto p(\yb|\xb,\thetab) \, p(\xb|\thetab) p(\thetab)
\propto p(\xb,\yb|\thetab) \, p(\thetab)
\propto p(\xb|\yb,\thetab) \, p(\thetab),
\]
The following  are some known schemes:\\
$\bullet$ Joint Maximum \aposteriori (JMAP):
\[
\left(\wh{\thetab}, \wh{\xb}\right)
 =\argmaxs{(\thetab,\xb)}{p(\xb,\thetab|\yb)}
\]
\[
\sido{\yb}{\large JMAP}{\wh{\xb}}{\wh{\thetab}}
\]
$\bullet$ Generalized Maximum Likelihood (GML):
\[
\left\{\begin{array}{l}
\disp{ \wh{\xb}^{(k)}
=\argmaxs{\xb}{ p(\xb|\yb;\thetab^{(k-1)}) } } \\
\disp{ \wh{\thetab}^{(k)}
=\argmaxs{\thetab}{p(\wh{\xb}^{(k)}|\yb,\thetab) p(\thetab)} }
\end{array}\right.
\]
\[
\dido{\yb}{\wh{\thetab}^0}{\large GML}{\wh{\xb}^{k}}{\wh{\thetab}^{k}}
\]
$\bullet$ Marginalized Maximum Likelihood (MML):
\[
\left\{\begin{array}{l}
\disp{ \wh{\thetab}=\argmaxs{\thetab}{
      \int p(\yb|\xb) \, p(\xb;\thetab) \d{\xb}}           } \\
\disp{ \wh{\xb}=\argmaxs{\xb}{p(\xb|\yb;\wh{\thetab})} }
\end{array}\right.
\]

\[
\siso{\yb}{\large ML}{\wh{\thetab}} \disou{}{\yb}{\large MAP}{\wh{\xb}}
\]
$\bullet$ MML-EM: \\
An analytic expression for $p(\yb;\thetab)$ is rarely possible.
Consequently, considering $[\yb, \xb]$ as the complete data and $\yb$ as
the incomplete data, we can use the EM algorithm to obtain the
following scheme:
\[
\left\{\begin{array}{l}
\hbox{E:~}
 Q\left(\thetab;\wh{\thetab}^{(k)}\right)
=\espx{\xb|\yb;\thetab^{(k)}}{\ln p(\xb,\yb;\thetab)} \\
\hbox{M:~}
 \wh{\thetab}^{(k+1)}
 =\disp{ \argmaxs{\thetab}{Q\left(\thetab;\wh{\thetab}^{(k)}\right)} }
\end{array}
\right.
\]
\[
\disod{\yb}{\wh{\thetab}^{(0)}}{ML-EM}{\wh{\thetab}^{(k)}}
\disod{\wh{\thetab}}{\yb}{MAP}{\wh{\xb}}
\]

\section{Data fusion}

In this section we consider some simple data fusion problems
and analysis the way we can use the previous schemes to solve them.

\subsection{Sensors without noise}

\medskip\noindent{\em Problem P6:}
The sensor C1 gives $N$ samples $\xb_a=\{x_1,\ldots, x_N\}$ of $X$
and stops.
The sensor C2 gives $M$ samples $\yb_b=\{y_1,\ldots, y_M\}$
related to $\xb$ by $\yb=\Ab\xb+\bb$. \\
We are asked to predict the unobserved samples
$\xb_b=\{x_{N+1},\ldots, x_{N+M}\}$ of $X$.
\[
\begin{array}{lccl}
         & \xb_a              & \xb_b                      \\
C1:    & x_1, \ldots , x_N & \ldots  \mbox{?} \ldots   \\
C2:    & \ldots             & y_1, \ldots , y_M          \\
         & \yb_a              & \yb_b
\end{array}\quad
\quad \dido{\xb_a}{\yb_b}{ Fusion ? }{\wh{\xb}_b}{\wh{\thetab}}
\]
We can propose the following solutions:
\bit
\item Use $\xb_a$ to estimate $\thetab$, the parameters of $p(\xb;\thetab)$
and use it then to estimate $\xb_b$ from $\yb_b$:
\[
\barr{l}
\wh{\thetab}=\disp{\argmaxs{\thetab}{L_a(\thetab)=\ln p(\xb_a;\thetab)}} \\
\wh{\xb}_b =\disp{\argmaxs{\xb_b}{p\left(\xb_b | \yb_b;\wh{\thetab}\right)}}
\earr
\quad
\siso{\xb_a}{\large ML}{}
\disou{\wh{\thetab}}{\yb_b}{\large MAP}{\wh{\xb}_b}
\]

\item Use both $\xb_a$ and $\yb_b$ to estimate $\xb_b$:
\[
\barr{rl}
\wh{\thetab}&=\disp{\argmaxs{\thetab}{L_a(p(\xb_a;\thetab)}} \\
(\wh{\xb}_b,\wh{\thetab})
&=\disp{\argmaxs{(\xb_b,\wh{\thetab})}{p\left(\xb_b,\thetab|\xb_a,\yb_b\right)}}
\earr
\]
\[
\siso{\xb_a}{\large ML}{}
\tido{\yb_b}{\wh{\thetab}}{\xb_a}{JMAP,GML or ML-EM}{\wh{\xb}_b}{\wh{\thetab}}
\]
\eit

\subsection{Fusion of homogeneous data}

\medskip\noindent{\em Problem P7:}
We have two types of data on the same unknown $\xb$, both related to it through
linear models:

\vspace{-9pt}
\bcc
\begin{tabular}{l}
\syslins{\xb}{\Hb_1}{\bb_1}{\yb} \\
\syslins{\xb}{\Hb_2}{\bb_2}{\zb}
\end{tabular}
\ecc
For example, consider an X ray tomography problem where
$\xb$ represents the mass density of the object and where $\yb$ and $\zb$
represent respectively a high resolution projection and a low resolution projection.

We can use directly the Bayesian approach to solve this problem:
\[
p(\xb|\yb,\zb)
=\frac{p(\yb,\zb|\xb) \, p(\xb)}{p(\yb,\zb)}
\]
Actually the main difficulty here is to assign $p(\yb,\zb|\xb)$.
If we assume that the errors associated to the two sets of data are
independent then the calculation can be done more easily.
For the purpose of illustration assume the following:
\beqnn
p(\yb|\xb;\sigma_1^2) &\propto& \expf{-\frac{1}{2\sigma_1^2} |\yb-\Hb_1 \xb|^2}
\\
p(\zb|\xb;\sigma_2^2) &\propto& \expf{-\frac{1}{2\sigma_2^2} |\zb-\Hb_2 \xb|^2}
\\
p(\xb;\mb,\Sigmab) &\propto&
\expf{-\frac{1}{2}
        [\xb-\mb]^t \Sigmab^{-1} [\xb-\mb]
     }
\eeqnn
Indeed, assume that the hyper-parameters $(\sigma_1^2,\sigma_2^2,\mb,\Sigmab)$
are given. Then we can use, for example, the MAP estimate, given by:
\[
\wh{\xb}=\argmaxs{\xb}{p(\xb|\yb,\zb)}=\argmins{\xb}{J(\xb)=J_1(\xb)+J_2(\xb)+J_3(\xb)}
\]
with
\[
J_1(\xb)= \frac{1}{2\sigma_1^2} |\yb-\Hb_1 \xb|^2,
\]
\[
J_2(\xb)= \frac{1}{2\sigma_2^2} |\zb-\Hb_2 \xb|^2,
\]
\[
J_3(\xb)= \frac{1}{2} [\xb-\mb]^t \Sigmab^{-1} [\xb-\mb]
\]
However, in practical applications, the data come from different processes.

\subsection{Real data fusion problems}

Consider a more realistic data fusion problem, where we have two different
kinds of data. As an example assume a tomographic image reconstruction
problem where we have a set of data $\yb$ obtained by an X ray and a set
of data $\zb$ obtained by an ultrasound probing system.
The X ray data are related to the mass density $\xb$ of the matter
while the ultrasound data are related to
the acoustic reflectivity $\rb$ of the matter.
Indeed, assume that, we have linear
relations, both between $\yb$ and $\xb$ and between $\zb$ and $\rb$.
Then we have:
\bcc
\begin{tabular}{l}
\syslins{\xb}{\Hb_1}{\bb_1}{\yb} \\
\syslins{\rb}{\Hb_2}{\bb_2}{\zb}
\end{tabular}
\ecc
Assuming that the two sets of data are independant, we can again use
the Bayes rule which now becomes
\[
p(\xb,\rb|\yb,\zb)
=\frac{p(\yb,\zb|\xb,\rb) \, p(\xb,\rb)}{p(\yb,\zb)}
=\frac{p(\yb|\xb) \, p(\zb|\rb) \, p(\xb,\rb)}{p(\yb,\zb)}
\]
with
\[
 p(\yb,\zb)= \intg\intg p(\yb|\xb) \, p(\zb|\rb) \, p(\xb,\rb) \d{\rb}\d{\xb}.
\]
Here also the main difficulty is the assignment of the probability laws
$p(\yb|\xb)$, $p(\zb|\rb)$, and more specifically $p(\xb,\rb)$.

Actually if we could find a mathematical relation between $\rb$ and $\xb$,
then the problem would become the same as in the preceding case.
To see this, assume that we can find a relation such as
$r_j=g(x_{i+1} -x_{i})$ with $g$ a monotonic increasing function,
from some physical reasons.
For example, the fact that in the area where there are some important
changes in the mass density of the matter both $\xb$ and $\rb$ change.
Indeed, if $g$ could be a linear function (an unrealistic hypothesis) then
we would have
\[
\left\{\barr{l}
\yb=\Hb_1\xb+\bb_1 \\
\zb=\Hb_2\rb+\bb_2 \\
\rb=\Gb\xb
\earr\right.
\lra
\left\{\barr{l}
\yb=\Hb_1\xb+\bb_1 \\
\zb=\Gb\Hb_2\rb+\bb_2
\earr\right.
\]

For more realistic cases we need a method which does not use a physically based
explicit expression of $g$.
One approach proposed and used by {\em Gautier et al.}
\cite{Gautier95b,Gautier95c,Gautier96,Gautier97a} is based on a
compound Markovian model where the body object $\ob$ is assumed to be
composed of three related quantities:
\[
\ob=\{\rb,\xb\}=\{\qb,\ab,\xb\}
\]
where $\qb$ is a binary vector representing the positions of the
discontinuities (edges) in the body, $\ab$ a vector containing the
reflectivity values such that
\[
 \left\{\barr{l} q_j=0 \lra r_j=0,\\ q_j=1 \lra r_j=a_j\earr\right.
\hbox{~~and~~}
r_j=\left\{\begin{array}{ll}
g(x_{j+1} -x_{j}) & \hbox{if~~}  |x_{j+1} -x_{j}|>\alpha \\
0                    & \hbox{otherwise}
\end{array}\right.
\]
and $g$ is any monotonic increasing function.

With this model we can write
\[
p(\ob,\rb)=p(\xb,\ab,\qb)=p(\xb|\ab,\qb) \, p(\ab|\qb) \, p(\qb)
\]
and using the Bayes rule, we have
\[
p(\xb,\ab,\qb|\yb,\zb)
\propto p(\yb,\zb|\xb,\ab,\qb)\, p(\xb,\ab,\qb)
=p(\yb,\zb|\xb,\ab,\qb) \, p(\xb|\ab,\qb) \, p(\ab|\qb) \, p(\qb)
\]
We illustrate this approach by making the following assumptions:

\bit
\item Conditional independence of $\yb$ and $\zb$:
$p(\yb,\zb|\xb,\ab,\qb)=p(\yb|\xb)p(\zb|\ab)$

\item Gaussian laws for $\bb_1$ and $\bb_2$
\[
p(\yb|\xb;\sigma_1^2)\propto \expf{-\frac{1}{2\sigma_1^2} |\yb-\Hb_1 \xb|^2};
\quad
p(\zb|\ab;\sigma_2^2)\propto \expf{-\frac{1}{2\sigma_2^2} |\zb-\Hb_2 \ab|^2}
\]

\item Bernoulli law for $\qb$: \quad
\(\disp{p(\qb) \propto \sum_{i=1}^n q_i^{\lambda} (1-q_i)^{1-\lambda}}\)

\item Gaussian law for $\ab|\qb$:
\[
p(\ab|\qb) \propto \expf{-\frac{1}{2\sigma_a^2} \ab^t \Qb \ab}, \quad
\Qb=\hbox{diag}[q_1,\ldots,q_n]
\]

\item Markovian model for $\xb$: \quad
\(
p(\xb|\ab,\qb) \propto \expf{-U(\xb|\ab,\qb)}
\)
\eit
Then, based on
\[
p(\xb,\ab,\qb|\yb,\zb)
 \propto p(\yb|\xb) \, p(\zb|\ab) \, p(\xb|\ab,\qb) \, p(\ab|\qb) \, p(\qb)
\]
we can propose the following schemes:
\bit
\item Simultaneous estimation of all the unknowns with the
joint MAP estimation (JMAP):
\[
\left(\wh{\xb},\wh{\ab},\wh{\qb}\right)
=\argmaxs{(\xb,\ab,\qb)}{p(\xb,\ab,\qb|\yb,\zb)}
\qquad
\dito{\yb}{\zb}{\large JMAP}{\wh{\xb}}{\wh{\ab}}{\wh{\qb}}
\]
\item First estimate the positions of the discontinuities $\qb$
and then use them to estimate $\xb$ and $\ab$ :
\[
\left\{\barr{l}
 \wh{\qb}=\disp{\argmaxs{\qb}{p(\qb|\yb,\zb)}} \\
\left(\wh{\xb},\wh{\ab}\right)
=\disp{\argmaxs{(\xb,\ab)}{p(\xb,\ab|\yb,\zb,\wh{\qb})}}
\earr\right. \quad
\diso{\yb}{\zb}{\hbox{Det.}}{}
\tido{\yb}{\wh{\qb}}{\zb}{\hbox{Est.}}{\wh{\xb}}{\wh{\ab}}
\]
\item First estimate the positions of the discontinuities $\qb$ using only
$\zb$ and then use them to estimate $\xb$ and $\ab$ :
\[
\left\{\barr{l}
 \wh{\qb}=\disp{\argmaxs{\qb}{p(\qb|\zb)}} \\
\left(\wh{\xb},\wh{\ab}\right)
=\disp{\argmaxs{(\xb,\ab)}{p(\xb,\ab|\yb,\zb,\wh{\qb})}}
\earr\right. \quad
\siso{\zb}{\hbox{Det.}}{}
\tido{\yb}{\wh{\qb}}{\zb}{\hbox{Est.}}{\wh{\xb}}{\wh{\ab}}
\]
\item First estimate $\qb$ and $\ab$ using only
$\zb$ and then use them to estimate $\xb$:

\[
\left\{\barr{l}
 \left(\wh{\qb},\wh{\ab}\right) = \disp{\argmaxs{\qb,\ab}{p(\qb,\ab|\zb)}} \\
 \wh{\xb} = \disp{\argmaxs{\xb}{p(\xb|\yb,\wh{\ab},\wh{\qb})}}
\earr\right. \quad
\sidotpup{\zb}{Det.}{\&}{Est.}{}{}
\hspace{-2mm}
\tiso{\wh{\qb}}{\wh{\ab}}{\yb}{Est.}{\wh{\xb}}
\]
\item First estimate only $\qb$ using $\zb$, then estimate $\ab$
using $\wh{\qb}$ and $\zb$, and finally, estimate $\xb$ using $\wh{\qb}$,
$\wh{\ab}$ and $\yb$:
\[
\left\{\barr{l}
 \disp{\wh{\qb}=\argmaxs{\qb}{p(\qb|\zb)}} \\
 \disp{\wh{\ab}=\argmaxs{\ab}{p(\ab|\zb,\wh{\qb})}} \\
 \disp{\wh{\xb}=\argmaxs{\xb}{p(\xb|\yb,\wh{\ab},\wh{\qb})}}
\earr\right.
\]
\[
\siso{\zb}{\hbox{Det.}}{\wh{\qb}}\quad
\diso{\wh{\qb}}{\zb}{\hbox{Est.}}{\wh{\ab}} \quad
\tiso{\wh{\qb}}{\wh{\ab}}{\yb}{\hbox{Est.}}{\wh{\xb}}
\]
\item First estimate only $\qb$ using $\zb$ and then estimate
$\xb$ using $\wh{\qb}$ and the data $\yb$:
\[
\left\{\barr{ll}
 \wh{\qb} &= \disp{\argmaxs{\qb}{p(\qb|\zb)}} \\
 \wh{\xb} &= \disp{\argmaxs{\xb}{p(\xb|\yb,\wh{\qb})}}
\earr\right.
\barr{l}\siso{\zb}{Det.}{\wh{\qb}}\\ ~\earr
\diso{\wh{\qb}}{\yb}{Est.}{\wh{\xb}}
\]
\eit
Two more realistic solutions are:

\bigskip\noindent{\em Proposed method 1:} \\
Estimate $\rb$ using only $\zb$ and estimate $\xb$ and $\qb$ using
$\wh{\rb}$ and $\yb$:
\[
\barr{ccc}
\siso{\zb}{\hbox{Est.}}{\wh{\rb}}
& ~\qquad~ &
\dido{\wh{\rb}}{\yb}{\hbox{Reconstruction}}{\wh{\xb}}{\wh{\qb}}
\\ ~\\
p(\rb|\zb)\propto p(\zb|\rb) \, p(\rb)
& ~\qquad~ &
p(\xb,\qb|\wh{\rb},\yb)\propto p(\yb|\xb) \, p(\xb,\qb|\wh{\rb})
\earr
\]
For the first part, with the assumptions made, we have
\[
 \wh{\rb} = \argmaxs{\rb}{p(\rb|\zb)}
          =\argmins{\rb}{J_1(\rb|\zb)}
\]
with
\[
J_1(\rb|\zb)
= |\zb-\Hb_2 \rb|^2 +\lambda \sum_j (r_{j+1}-r_j)^2
\]
and for the second part we have
\[
\left(\wh{\xb},\wh{\qb}\right) = \argmaxs{(\xb,\qb)}{p(\xb,\qb|\yb,\wh{\rb})}
                =\argmins{(\xb,\qb)}{J_2(\xb,\qb|\yb,\wh{\rb})} \\
\]
with
\[
J_2(\xb,\qb|\yb,\wh{\rb})
= |\yb-\Hb_1 \xb|^2+\lambda \sum_j (1-q_j) (x_{j+1}-x_j)^2
+\alpha_1 \sum_j q_j (1-\wh{r}_j)+\alpha_2 \sum_j q_j \wh{r}_j
\]
This last optimization is still too difficult to do due to the presence of
$\qb$ and $\xb$ together. An easier solution is given below.

\medskip\noindent{\em Proposed method 2:} \\
Use the ultrasound data $\zb$ to detect the locations of some of
the boundaries and use X ray data to make an intensity image preserving
the positions of these discontinuities:

\[
\siso{\zb}{Est.}{\wh{\rb}}
\siso{}{$q_j=\frac{|r_j|}{\sum_j |r_j|}$}{\wh{\qb}}
\quad
\diso{\wh{\qb}}{\yb}{Est.}{\wh{\xb}}
\]
Here, we made slightly different assumptions about the distributions of
$\rb$ and $\xb$. Actually a generalized Gaussian distribution in
place of Gaussian gives a good compromise of discontinuity preservation and
easy implementation. A typical choice, for the first case is
\beqnn
 \wh{\rb} &=& \argmaxs{\rb}{p(\rb|\zb)}
            =\argmins{\rb}{J_1(\rb|\zb)} \\
\hbox{with}\qquad
J_1(\rb|\zb)
&=& ||\zb-\Hb_2 \rb||^2 +\lambda_1 ||\rb||^p, \quad 1<p<2
\eeqnn
and for the second case is
\beqnn
\wh{\xb} &=& \argmaxs{\xb}{p(\xb|\yb,\wh{\qb})}
            = \argmins{\xb}{J_2(\xb|\yb;\wh{\qb})} \\
\hbox{with}\qquad
J_2(\xb|\yb,\wh{\qb})
&=& ||\yb-\Hb_1 \xb||^2 + \lambda_2 \sum_j (1-q_j) |x_{j+1}-x_j|^p,
\quad 1<p<2
\eeqnn
The aim of this paper is not to go through more details on these methods.
The interested reader should refer to \cite{Gautier97a,Gautier97b}.

\section{Conclusions}
To conclude briefly:
\bit
\item ME can be used when we want to assign a probability law $p(\xb)$
from some expected values.

\item ML can be used when we have a parametric form of the probability law
$p(\xb,\thetab)$ and we have access to direct observations $\xb$ of $X$,
and we want to estimate the parameters $\thetab$.

\item ML-EM extends the ML to the case of incomplete observations.

\item When the observed data are noisy the Bayesian approach is the
most appropriate.

\item For practical data fusion problems the Bayesian approach seems to give
all the necessary tools we need.

\item Compound Markov models are convenient models to represent
signals and images in a Bayesian approach of data fusion.

\item The Bayesian approach is coherent and easy to understand.
However, in real applications, we have still much to do to implement it: \\
-- Assignment or choice of the prior laws \\
-- Efficient optimization of the obtained criteria \\
-- Estimation of the hyper-parameters \\
-- Interpretation of the obtained results.
\eit

%
%

\def\AsAs{Astrononmy and Astrophysics}					
\def\AAP{Advances in Applied Probability}				
\def\ABE{Annals of Biomedical Engineering}				
\def\AISM{Annals of Institute of Statistical Mathematics}		
\def\AMS{Annals of Mathematical Statistics}			
\def\AO{Applied Optics}							
\def\AP{The Annals of Probability}					
\def\ARAA{Annual Review of Astronomy and Astrophysics}			
\def\AST{The Annals of Statistics}					
\def\AT{Annales des T\'el\'ecommunications}				
\def\BMC{Biometrics}							
\def\BMK{Biometrika}							
\def\CPAM{Communications on Pure and Applied Mathematics}		
\def\EMK{Econometrica}							
\def\CRAS{Compte-rendus de l'acad\'emie des sciences}			
\def\CVGIP{Computer Vision and Graphics and Image Processing}		
\def\GJRAS{Geophysical Journal of the Royal Astrononomical Society}	
\def\GSC{Geoscience}						
\def\GPH{Geophysics}							
\def\GRETSI#1{Actes du #1$^{\mbox{e}}$ Colloque GRETSI} 		
\def\CGIP{Computer Graphics and Image Processing}			
\def\ICASSP{Proceedings of IEEE ICASSP}					
\def\ICEMBS{Proceedings of IEEE EMBS}					
\def\ICIP{Proceedings of the International Conference on Image Processing}
\def\ieeP{Proceedings of the IEE}					
\def\ieeeAC{IEEE Transactions on Automatic and Control}			
\def\ieeeAES{IEEE Transactions on Aerospace and Electronic Systems}	
\def\ieeeAP{IEEE Transactions on Antennas and Propagation}		
\def\ieeeAPm{IEEE Antennas and Propagation Magazine}			
\def\ieeeASSP{IEEE Transactions on Acoustics Speech and Signal Processing}
\def\ieeeBME{IEEE Transactions on Biomedical Engineering}		
\def\ieeeCS{IEEE Transactions on Circuits and Systems}			
\def\ieeeCT{IEEE Transactions on Circuit Theory}			
\def\ieeeC{IEEE Transactions on Communications}				
\def\ieeeGE{IEEE Transactions on Geoscience and Remote Sensing}		
\def\ieeeGEE{IEEE Transactions on Geosciences Electronics}		
\def\ieeeIP{IEEE Transactions on Image Processing}			
\def\ieeeIT{IEEE Transactions on Information Theory}			
\def\ieeeMI{IEEE Transactions on Medical Imaging}			
\def\ieeeMTT{IEEE Transactions on Microwave Theory and Technology}	
\def\ieeeM{IEEE Transactions on Magnetics}				
\def\ieeeNS{IEEE Transactions on Nuclear Sciences}			
\def\ieeePAMI{IEEE Transactions on Pattern Analysis and Machine Intelligence}
\def\ieeeP{Proceedings of the IEEE}					
\def\ieeeRS{IEEE Transactions on Radio Science}				
\def\ieeeSMC{IEEE Transactions on Systems, Man and Cybernetics}		
\def\ieeeSP{IEEE Transactions on Signal Processing}			
\def\ieeeSSC{IEEE Transactions on Systems Science and Cybernetics}	
\def\ieeeSU{IEEE Transactions on Sonics and Ultrasonics}		
\def\ieeeUFFC{IEEE Transactions on Ultrasonics Ferroelectrics and Frequency Control}
\def\IJC{International Journal of Control}				
\def\IJCV{International Journal of Computer Vision}			
\def\IJIST{International Journal of Imaging Systems and Technology}	
\def\IP{Inverse Problems}						
\def\ISR{International Statistical Review}				
\def\IUSS{Proceedings of International Ultrasonics Symposium}		
\def\JAPH{Journal of Applied Physics}					
\def\JAP{Journal of Applied Probability}				
\def\JAS{Journal of Applied Statistics}					
\def\JASA{Journal of Acoustical Society America}			
\def\JASAS{Journal of American Statistical Association}			
\def\JBME{Journal of Biomedical Engineering}				
\def\JCAM{Journal of Computational and Applied Mathematics}		
\def\JEWA{Journal of Electromagnetic Waves and Applications}		
\def\JMO{Journal of Modern Optics}					
\def\JNDE{Journal of Nondestructive Evaluation}				
\def\JMP{Journal of Mathematical Physics}				
\def\JOSA{Journal of the Optical Society of America}			
\def\JP{Journal de Physique}						
\def\JRSSA{Journal of the Royal Statistical Society A}			
\def\JRSSB{Journal of the Royal Statistical Society B}			
\def\JRSSC{Journal of the Royal Statistical Society C}			
\def\JSPI{Journal of Statistical Planning and Inference}  		
\def\JTSA{Journal of Time Series Analysis}                   		
\def\JVCIR{Journal of Visual Communication and Image Representation} 	
	\def\MMAS{???} 
\def\KAP{Kluwer \uppercase{A}cademic \uppercase{P}ublishers}							%
\def\MNAS{Mathematical Methods in Applied Science}			
\def\MNRAS{Monthly Notices of the Royal Astronomical Society}		
\def\MP{Mathematical Programming}					
	\def\NSIP{NSIP}  
\def\OC{Optics Communication}						
\def\PRA{Physical Review A}						
\def\PRB{Physical Review B}						
\def\PRC{Physical Review C}						
\def\PRD{Physical Review D}						
\def\PRL{Physical Review Letters}					
\def\RGSP{Review of Geophysics and Space Physics}			
\def\RPA{Revue de Physique Appliqu\'e}							
\def\RS{Radio Science}							
\def\SP{Signal Processing}						
\def\siamAM{SIAM Journal of Applied Mathematics}			
\def\siamCO{SIAM Journal of Control}					
\def\siamJO{SIAM Journal of Optimization}				
\def\siamMA{SIAM Journal of Mathematical Analysis}			
\def\siamNA{SIAM Journal of Numerical Analysis}				
\def\siamR{SIAM Review}							
\def\SSR{Stochastics and Stochastics Reports}       			
\def\TPA{Theory of Probability and its Applications}			
\def\TMK{Technometrics}							
\def\TS{Traitement du Signal}						
\def\UCMMP{U.S.S.R. Computational Mathematics and Mathematical Physics}	
\def\UMB{Ultrasound in Medecine and Biology}				
\def\US{Ultrasonics}							
\def\USI{Ultrasonic Imaging}						

%
\def\jan{janvier\xspace}
\def\feb{f\'evrier\xspace}
\def\mar{mars\xspace}
\def\apr{avril\xspace}
\def\may{mai\xspace}
\def\jun{juin\xspace}
\def\jul{juillet\xspace}
\def\aug{ao\^ut\xspace}
\def\sep{septembre\xspace}
\def\oct{octobre\xspace}
\def\nov{novembre\xspace}
\def\dec{d\'ecembre\xspace}
\def\Jan{January\xspace}	
\def\Feb{February\xspace}
\def\Mar{March\xspace}
\def\Apr{April\xspace}
\def\May{May\xspace}
\def\Jun{June\xspace}
\def\Jul{July\xspace}
\def\Aug{August\xspace}
\def\Sep{September\xspace}
\def\Oct{October\xspace}
\def\Nov{November\xspace}
\def\Dec{December\xspace}
\def\sub{soumis \`a\xspace}
 \def\UP#1{\uppercase{#1}}
 \bibliographystyle{ieeetr}

\edoc